\documentclass[amssymb,amsfonts,twocolumn,floatfix,showpacs,superscriptaddress,10pt]{revtex4-1}

\usepackage[utf8]{inputenc}
\usepackage[french,english]{babel}
\usepackage[T1]{fontenc}
\usepackage[version=3]{mhchem}
\usepackage{graphicx}
\usepackage{dcolumn}
\usepackage{bm}
\usepackage{xcolor}
\usepackage{subcaption}
\usepackage{amssymb}
\usepackage{amsmath}
\usepackage[amsmath]{ntheorem}
\usepackage{epsfig}
\usepackage[colorlinks=false, pdfborder={0 0 0}]{hyperref}
\usepackage{cleveref}

\crefname{figure}{Figure}{Figures}
\crefname{equation}{Equation}{Equations}
\crefname{table}{Table}{Tables}

\begin{document}
\title{Optimal laser pulse energy partitioning for air ionization}

\author{Elise Schubert}
\affiliation{Universit\'e de Gen\`eve, GAP, 22 chemin de Pinchat, 1211 Gen\`eve 4, Switzerland}
\author{Jean-Gabriel Brisset}
\affiliation{Universit\'e de Gen\`eve, GAP, 22 chemin de Pinchat, 1211 Gen\`eve 4, Switzerland}
\affiliation{Max-Born Institut, Max-Born-Strasse 2A, 12489 Berlin, Germany}
\author{Mary Matthews}
\affiliation{Universit\'e de Gen\`eve, GAP, 22 chemin de Pinchat, 1211 Gen\`eve 4, Switzerland}
\author{Antoine Courjaud}
\affiliation{Amplitude Syst\`emes, 11, avenue de Canteranne, Cit\'e de la Photonique, 33600 Pessac, France}
\author{J\'er\^ome Kasparian}\email{jerome.kasparian@unige.ch}
\affiliation{Universit\'e de Gen\`eve, GAP, 22 chemin de Pinchat, 1211 Gen\`eve 4, Switzerland}
\author{Jean-Pierre Wolf}
\affiliation{Universit\'e de Gen\`eve, GAP, 22 chemin de Pinchat, 1211 Gen\`eve 4, Switzerland}

\begin{abstract}
We investigate the pulse partitioning of a 6.3~mJ, 450~fs pulse at 1030~nm to produce plasma channels. At such moderate energies, splitting the energy into several sub-pulses reduces the ionization efficiency and thus does not extend the plasma lifetime. We numerically show that when sufficient energy to produce multifilamentation is available, splitting the pulse temporally in a pulse train increases the gas temperature compared to a filament bundle of the same energy. This could improve the mean free path of the free electrons, therefore enhancing the efficiency of discharge triggering.
\end{abstract}

%\pacs{42.65.Jx Beam trapping, self-focusing and
%defocusing; self-phase modulation, 42.65.Tg Optical solitons; nonlinear guided
%waves, 32.80.Fb Photoionization of atoms and ions, Laser-produced plasma, 52.50.Jm}

\maketitle

\section{Introduction}

Laser-induced ionization is a key step in many applications of ultrashort laser pulses, including laser ablation \cite{miller1998laser,Mathis2012}, laser-induced breakdown spectroscopy (LIBS) \cite{cremers2000laser,StelmRMYSKAWW2004}, fast electric switching \cite{Kirkman1986,Tkotz1995}, or the control of high-voltage discharges \cite{ZhaoDWE1995,LaCCDGJJKMMPRVPCM2000,RodriSWWFAMKRKKSYW2002} and lightning \cite{KaspaAAMMPRSSYMSWW2008a}.

Optimizing the efficiency of these processes requires to maximize both the ionization yield and the free electron lifetime which can be improved by heating the surrounding gas. Using sequences of multiple pulses have been shown to be relevant in this purpose, in the case of LIBS \cite{Sattmann1995}, laser ablation in solids \cite{Chichkov1996,Vorobyev2005} or dielectrics \cite{Jasapara2001}, and filamentation.

Such effort is especially relevant in the case of laser filamentation, a non-linear propagation regime typical of high-power ultrashort laser pulses~\cite{ChinHLLTABKKS2005,CouaiM2007,BergeSNKW2007}. An interplay between focusing by the Kerr effect and defocusing effects including ionization and the saturation of the Kerr effect \cite{BejotKHLVHFLW2010a,bejot2011transition} ensures intensities in the range of several tens of TW/cm$^2$ over long distances \cite{LaVJCCDJKP1999,RodriBMKYSSSELHSWW2004}.

Multiple pulses can increase the ionization yield and free electron lifetime through different processes \cite{Wang2011}, including by triggering avalanche ionization \cite{Shen1984}, photo-detaching electrons from O$_2^-$ ions \cite{BurchSB1958} resulting from the electron attachment on oxygen molecules, or creating a plasma with a first pulse and heating it with subsequent pulse(s), by Joule heating (or inverse Bremestrahlung) enhancing the cascade ionization efficiency \cite{PolynM2011}. 

However, most results reported to date investigate the effect of adding one or several subsequent pulses to a main pulse. Such addition can only result in an increase of the observed effect. 
For example, launching a frequency-doubled YAG pulse of 7~ns duration at 532~nm together with a near-infrared (800~nm) pulse significantly increased the probability of triggering a discharge over 1.2~m,  lowering the median breakdown voltage by at least 5~\% \cite{M'ejAKSYWRKRSW2006,diels1992discharge,papeer2014extended,PolynM2011,zhou2009revival}. 
Alternatively, a 60-ns long sequence of over 30 pulses, with a total energy of 120~mJ, maintains the ionization over a duration comparable to the sequence itself, each pulse interacting more or less independently with the plasma~\cite{lu2015quasi}.

In contrast, in the present work, we consider a laser system with a fixed total energy and we aim at optimizing the ionization with an optimal temporal partitioning of the pulse energy. This is in particular characterized by the relative energy splitting and the delay(s) between sub-pulse(s). 
Such situation has been well investigated for nanosecond pulses with microsecond interpulse delays \cite{Sattmann1995}. In the case of ultrashort pulses, it has been only addressed in one particular situation: splitting a 30~mJ, 30 fs pulse into two sub-pulses separated by 7~ns decreased the ionization as characterized by acoustic signal, and had no impact on the pulse ability to guide electric discharges~\cite{ZhangLLHZWLZ2009}.

Here, we investigate the effect on the ionization yield and the lifetime of the resulting plasma of splitting an ultrashort laser pulse of several mJ into a pulse train of two  
sub-pulses with arbitrary relative intensities and delays. Furthermore, based on numerical simulations, we show that at a higher energy allowing multiple filamentation, a sequence of filamenting sub-pulses increases the temperature of the heavy species, as compared to the bundle of multiple filaments with the same energy. This could improve the efficiency of discharge triggering.

\section{Material and methods}

\begin{figure*}[t]
\centering
\includegraphics[width=15cm, keepaspectratio]{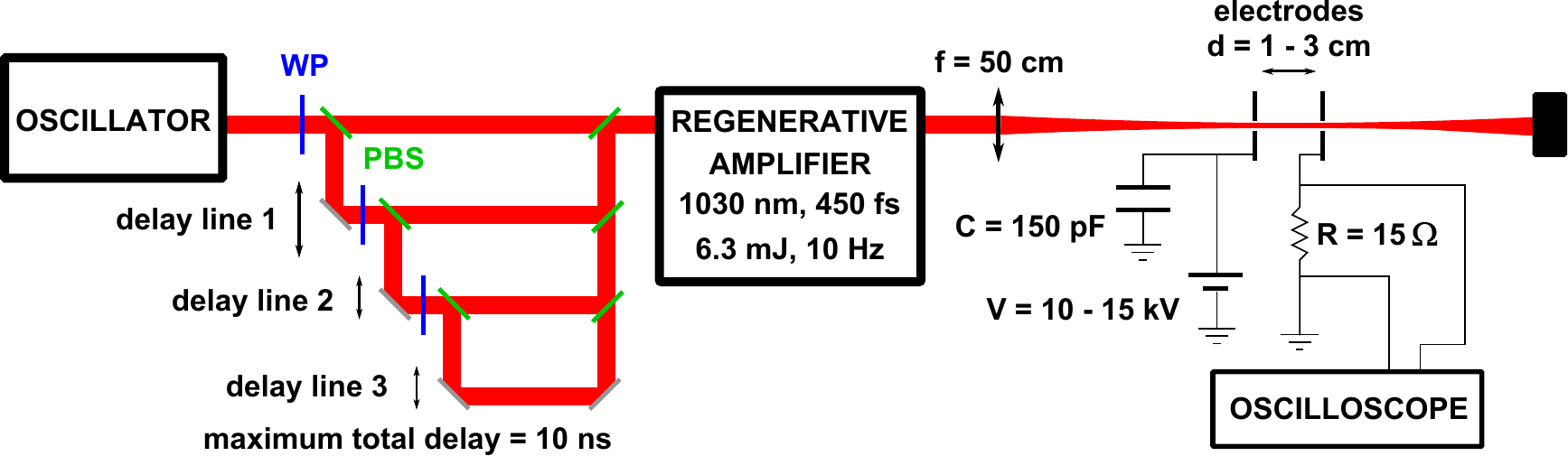} 
\caption{Experimental setup. blue WP: half-wave plate; green PBS: polarizing beam splitter.} 
\label{fig:Setup}
\end{figure*}

We used a chirped pulse amplifier (CPA) laser system based on a Yb:CaF$_2$ crystal as gain medium. The system delivers 450~fs FWHM pulses with a central wavelength of 1030~nm and a spectral bandwidth of 5~nm at a 10~Hz repetition rate. The total output energy of the system is  6.3~mJ. As sketched on~\cref{fig:Setup}, it can be split between one to four pulses with arbitrary amplitude ratios and tunable delays between the pulses over a range of 10~ns, limited by the length of the regenerative amplifier cavity. To achieve that purpose, the oscillator outcoming pulse is split into multiple pulses by a series of tunable passive beam splitters. The pulses are then recombined and amplified in the regenerative amplifier. The Yb:CaF$_2$ crystal has a saturation fluence of 73~$\rm J \cdot m^{-2}$ (as compared to 1~$\rm J \cdot m^{-2}$ for Ti:Sa) \cite[Chap.~11]{ricaud2012lasers,trager2007springer}. This very high saturation ratio allows the train of pulses to be amplified without modifying the relative amplitudes while extracting the maximum available energy. The output beam of the laser has a diameter of 6~mm and an $M^2$ of 1.22. The peak power obtained is 13.7~GW, above the 7~GW critical power \cite{KaspaSC2000} for filamentation at this wavelength~\cite{BergeSNKW2007}. To help self-focusing and to keep the focus at the same position independently of the energy of the individual pulses, a $f=$~50~cm lens is added after the laser amplifier. In this work, we focus on the effect of splitting the energy into two subpulses as the intensity was too weak for filamentation in the case of three and four subpulses.

To measure the effect of the laser filament on conductivity, two 2.5~cm square electrodes were placed in the beam, with holes allowing the beam to propagate through the center of the plates without touching the electrodes. One electrode is connected to a DC~HV power supply that provides up to 25~kV. A $C = 150$~pF capacitor is placed in parallel to the HV power supply to act as a charge reservoir. The other electrode is placed 1--3~cm apart from the first one, centered on the filamenting region and grounded. A $R = 15\ \Omega$ resistance is placed in series to probe the current flowing from the electrode to the ground. A DPO~7254C 2.5~GHz bandwidth oscilloscope is used to record the signal. The recording is triggered with a photodiode aiming at the beam dump after the electrodes and the data has been analysed using Matlab.

We checked that the time-integrated measured intensity, i.e. the total collected charge, is proportional to its maximum value. This ensures that the initial charge of the capacitor is only marginally depleted, so that our measurements are indeed representative of the air ionization in the plasma, and the measurements do not impact the measured plasma lifetime. In the following, we use the maximal value of the signal as a measurement of ionization.

The error bars were estimated from the variability of the electron yield with full laser power over a whole day. We attribute them mainly to the environmental changes in the laboratory (relative humidity and temperature), as well as laser fluctuations.

The free electron lifetime was defined as the longest time interval over which the electron density was kept over 1.6$\times 10^{20}$ m$^{-3}$. This threshold density is typical of that of a leader~\cite{ihaddadene2015increase}: the lifetime corresponds to the typical interval over which the plasma left behind by the laser would have a substantial influence on the atmospheric electricity behavior. One may note that this threshold is equivalent to 10\% of the peak value for a single pulse containing  the full 6.3~mJ energy of the laser. 

\textbf{Modeling.}
In order to better understand the observed results, we modeled the evolution of the plasma under multiple pulse illumination. The evolution of the electrons density is driven by~\cite{ZhaoDWE1995,fernsler1979nrl}: 

\begin{eqnarray}
\frac{\textrm{d}N_e}{\textrm{d}t}  
&=& R_{\textrm{col}} + R_{\textrm{av}} + R_{\textrm{ion}} \nonumber\\
&-& R_{\textrm{ep}} - R_{\textrm{at}} + R_{\textrm{pd}} \label{eq:Ne}
\end{eqnarray}
where
\begin{eqnarray}
R_{\textrm{col}} & = & \alpha N_e \label{eq:R1}\\ 
R_{\textrm{av}} & = & \frac{1}{\omega_0^2 \tau^2+1} \frac{q_e^2 \tau}{c \varepsilon_0 m_e U_{\textrm{p,O}_2}} I_L N_e \label{eq:R2}\\ 
R_{\textrm{ion}} & = & N_{\textrm{Mol}} W(I_L) \label{eq:R3}\\ 
R_{ \textrm{ep}} &=& \beta_{ep} T_e^{-0.7} N_e N_p \label{eq:R4}\\
R_{\textrm{at}} & = & (\eta_2 + \eta_3) N_e \label{eq:R5}\\ 
R_{\textrm{pd}} & = & \frac{N_n \sigma_{\textrm{O}_2^-} I_L}{\hbar \omega_0} \label{eq:R6}
\end{eqnarray}
where $W(I_L)$ 
describes the probability of ionization calculated
with the Perelomov, Popov, Terentev (PPT) formula~\cite{perelomov1966ionization} and $I_L$ is the incident laser intensity.  
The terms explicited in Equations~\eqref{eq:R1} to~\eqref{eq:R6} account for collision ionisation, avalanche ionization, probability of ionization, 
electron-ion recombination, attachment, and photodetachment, respectively.  
In the collision ionization term, $\tau = {1}/{\nu_{en}} = 1 / \left(10^{-13}  N_{\textrm{Mol}}  \sqrt{T_{e,[eV]}}\right)$ is the inverse of the electron-neutral molecule collision frequency. Two- and three-body electron attachment to oxygen molecules respectively occur at rates~\cite{ZhaoDWE1995}:
\begin{eqnarray}
\eta_2 &=& 1.22 \times 10^8 \frac{N_{\textrm{Mol}}}{N_0} \exp\left(-\frac{42.3}{E_0}\right) \label{eq:eta2}\\
\eta_3 &=& 10^8  \left(\frac{N_{\textrm{Mol}}}{N_0}\right)^2 \frac{ 0.62 + 800 E_0^2}{1 + 1000 E_0^2 \left[E_0 (1 + 0.03 E_0^2)\right]^{1/3}} \nonumber \\
&\times& \frac{T_{0,[eV]} e^{\frac{0.052}{T_{0,[eV]}}}}{ T_{e,[eV]} e^{\frac{0.052}{T_{e,[eV]}}}} \label{eq:eta3}
 \end{eqnarray}
 
The collision ionization rate $\alpha$ is given by~\cite{fernsler1979nrl}, displaying a very fast rise as a function of the temperature. All parameters are summarized in Table~\ref{tab:parameters}.

Similarly, the evolution of positive and negative ion densities, respectively, writes~\cite{ZhaoDWE1995,fernsler1979nrl}
\begin{eqnarray}
  \frac{\textrm{d}N_p}{\textrm{d}t} 
  &=& R_{\textrm{col}} + R_{\textrm{av}} + R_{\textrm{ion}} - R_{\textrm{ep}} \nonumber\\
  &-& \beta_{np} N_n N_p \left(\frac{T_g}{T_0}\right)^{-1.5} \label{eq:Np}
\end{eqnarray}

\begin{eqnarray}
 \frac{\textrm{d}N_n}{\textrm{d}t} 
 &=& R_{\textrm{at}} - R_{\textrm{pd}} - \beta_{np} N_n N_p \left(\frac{T_g}{T_0}\right)^{-1.5}  \label{eq:Nn}
\end{eqnarray}
The last term in Equations \eqref{eq:Np} and \eqref{eq:Nn} accounts for ion-ion recombination, where $T_g$ is the temperature of the heavy species and $T_0$ the ambient temperature. 

The total electron thermal energy evolves under Joule heating by the laser field and the external DC field, energy exchanges with the heavy species and the vibrational energy of air molecules, the excess energy in the ionization and photodetachment processes, losses of kinetic energy due to collision and avalanche ionization, electron losses related to electron-ion recombination, attachment, and transfer to molecules via impact excitation~\cite{fernsler1979nrl,papeer2014extended}: 
\begin{eqnarray}
\frac{\textrm{d}T_e}{\textrm{d}t}
	&=& \frac{2 J_L}{3 N_e  k_B} + \frac{2 q_e \mu_e E^2 }{3 k_B}  \nonumber\\
	&-& (T_e - T_v) \nu_{ev} - 2 (T_e - T_g)  \frac{m_e \nu_c}{M_{\textrm{air}}} \nonumber\\ 
	&+&   \left[R_{\textrm{ion}} U_e
	+ R_{pd} U_{e_{O_2^-}}  
	- \left(R_{\textrm{col}} + R_{\textrm{av}} \right) U_{\textrm{O}_2}  
	 \right] \cdot \frac{2}{3 k_B N_e} \nonumber\\ 
	&-& \left[ \frac{R_{\textrm{ep}}}{N_e} + (\eta_2 + \eta_3)  \right] T_e 
	- \frac{2 N_{\textrm{Tot}}}{3 k_B} R_{\textrm{imp}} 
\end{eqnarray}
where
\begin{eqnarray}
R_{\textrm{imp}} 
	& = & k_{\textrm{N}_2,\textrm{A}} U_{\textrm{N}_2,\textrm{A}} 
	+ k_{\textrm{N}_2,\textrm{B}} U_{\textrm{N}_2,\textrm{B}}  \nonumber\\
	& + & k_{\textrm{N}_2,\textrm{C}} U_{\textrm{N}_2,\textrm{C}} 
	+ k_{\textrm{O}_2,\textrm{a}} U_{\textrm{O}_2,\textrm{a}} 
	+ k_{\textrm{O}_2,\textrm{b}} U_{\textrm{O}_2,\textrm{b}}
\end{eqnarray}
and $\mu_e$ is the electron mobility \cite{ZhaoDWE1995}:
\begin{equation}
\mu_e (m^2/V\cdot s) = -\frac{N_0}{3N_\textrm{Tot}}\left(\frac{5\times10^5+E_0}{1.9\times10^4+26.7\times E_0}\right)^{0.6}
\label{eq:mobility}
\end{equation}
$N_0$ being the molecule density at 1 atm and ${E_0=E N_0 / N_\textrm{Tot}}$. 

The heating rate is given by~\cite{*[{}] [{ Note that an $\varepsilon_0$ is missing in the definition of the Ohmic heating $J_L$ on p.~7.}] papeer2014extended}:  
\begin{equation}
J_L = \frac{4 \pi q_e^2 N_e \nu_\textrm{ei}} {m_e  c  \varepsilon_0 \left(\omega_0^2 + \nu_\textrm{ei}^2\right)} I_L
\end{equation}

Similarly the vibration and kinetic temperatures of the heavy species evolve as~\cite{papeer2014extended}: 
\begin{eqnarray}
\frac{\textrm{d}E_v}{\textrm{d}t}   &=& \frac{3}{2} N_e k_B (T_e - T_v) \nu_{ev} - \frac{E_v - E_{v,0}}{\tau_{VT}} \\
\frac{\textrm{d}T_g}{\textrm{d}t}
	&=& 2 (T_e - T_g) \frac{m_e \nu_c N_e}{M_{\textrm{air}} N_{\textrm{Tot}}} 
	+ \frac{2 (E_v - E_{v,0})}{3 \tau_{VT} k_B N_{\textrm{Tot}}}
\end{eqnarray}
where the vibrational energy is given by~\cite{ShneiZM2011}: 
\begin{equation}
E_v = \frac{ N_{\textrm{Tot}} (\hbar \omega_{\textrm{vib,N}_2})} {\exp \left(\frac{\hbar \omega_{\textrm{vib,N}_2}}{k_B T_v} \right) - 1}  
\end{equation}
and $E_{v,0}$ is its value for $T_v = T_0$, where $T_0$ is the ambient temperature. 
The cooling frequency  $\nu_\textrm{ev}$ is given by ~\cite{papeer2014extended}: 
\begin{equation}
\nu_{ev} = \frac{Q_c N_{\textrm{Mol}}}{1.5 k_B T_e}
\end{equation}

The electron-heavy species collision rate is the sum of the electron collision rates with molecules and positive ions~\cite{papeer2014extended}: 
\begin{equation}
\nu_{c} = 10^{-13} N_{\textrm{Mol}} \sqrt{T_{e,[eV]}} + 10^{-11} N_p T_{e,[eV]}^{-1.5}
\label{eq:nu_c}
\end{equation}

Finally, the vibrational-translational relaxation time $\tau_{VT}$  is~\cite{ShneiZM2011}: 
\begin{eqnarray}
    \tau_{VT} 
    &=& \biggl \lbrack N_{\textrm{Mol}}\cdot\biggl(7\cdot 10^{-16}  \exp\biggl(-\frac{141}{T_g^{1/3}}\biggr) \nonumber \\
    &+& 0.21\cdot 5\cdot 10^{-18} \exp\biggl(-\frac{128}{T_g^{0.5} }\biggr)\biggr) \biggl \rbrack^{-1}
\end{eqnarray}
  
\begin{table*}
\caption{Notations and parameters of the model.}
\begin{center}
\begin{tabular}{|c|c|c|c|}
\hline
Symbol & Meaning & Value & Reference \\ \hline
$\alpha$ & Collision ionization coeff. &  & \cite{fernsler1979nrl} \\
$W(I_L)$ & Probability of ionization &  & \cite{perelomov1966ionization} \\
$k_B$ & Bolzmann's constant & $1.38 \times 10^{-23}$ J/K & \\
$\omega_0$ & Laser frequency & $2.3546 \times 10^{15}$ s$^{-1}$ ($\lambda = 1030\textrm{ nm}$) & \\
$I_L$ & Incident laser intensity & & \\
$T_0$ & Ambient temperature & 300 K & \\
$T_{0,\textrm{[eV]}}$ & Ambient temperature in eV & 0.026 eV & \\
$M_{\textrm{air}}$ & Average mass of air molecules & $4.7704 \times 10^{-26}$ kg & \\
$N_0$ & Density of molecules (normal conditions) & $2.688\times 10^{25}$ m$^{-3}$ & \\
$q_e$ & Electron charge & $1.6\times10^{-19}$ C & \\
$m_e$ & Electron mass & $9.1\times 10^{-31}$ kg & \\
$\sigma_{\textrm{O}_2^-}$ & Photodetachment cross section of O$_2^-$ & $3.8\times10^{-23}$ m$^2$ & \cite{BurchSB1958} \\ 
$\beta_{ep}$ & Electron-ion recombination coefficient & $1.138\times 10^{-11}$  K$^{0.7}$m$^3$/s & \cite{ZhaoDWE1995} \\
$\beta_{np}$ & Ion-ion recombination coefficient & $2.1792\times 10^{-13}$ m$^3$/s & \cite{ZhaoDWE1995} \\ 
$\eta_2$ & 2-body attachment coefficient & See Eq.~\eqref{eq:eta2} & \cite{ZhaoDWE1995,fernsler1979nrl}\\
$\eta_3$ & 3-body attachment coefficient & See Eq.~\eqref{eq:eta3}  & \cite{ZhaoDWE1995,fernsler1979nrl}\\ 
$\hbar \omega_{\textrm{vib,N}_2}$ & Quantum of vibration of N$_2$ & 0.29 eV ($4.64 \times 10^{-20}$ J) & \cite{ShneiZM2010} \\ 
$U_{\textrm{N}_2,\textrm{A}}$ & Energy of the N$_2$(A$^3_\Sigma$) state & 6.2 eV & \cite[p.~13]{fernsler1979nrl}\\ 
$U_{\textrm{N}_2,\textrm{B}}$ & Energy of the N$_2$(B$^3_\pi$) state & 7.35 eV & \cite{valdivia1997physics} \\ 
$U_{\textrm{N}_2,\textrm{C}}$ & Energy of the N$_2$(C$^3_\pi$) state & 11 eV & \cite{cartwright1977electron}\\ 
$U_{\textrm{O}_2,\textrm{a}}$ & Energy of the O$_2$(a$^1_\Delta$) state & 1 eV & \cite[p.~14]{fernsler1979nrl}\\ 
$U_{\textrm{O}_2,\textrm{b}}$ & Energy of the O$_2$(b$^1_\Sigma$) state & 1.6 eV & \cite[p.~14]{fernsler1979nrl}\\ 
$k_\textrm{X}$ & Rate of electron impact excitation to state $X$ & \cite[Table~II]{fernsler1979nrl} 
& \cite[p.~14]{fernsler1979nrl}\\ 
$E$ &  External DC field & 10-14~kV & \\
$\mu_e$ &  Electron mobility & See Eq.~\eqref{eq:mobility} & \cite{ZhaoDWE1995}\\
%$N_0$ &  Density of air at 1 atm & $2.5 \times 10^{25}$~m$^{-3}$ & \\
$N_{\textrm{Tot}}$ &  Density of heavy species (initial density of molecules) & $2.5 \times 10^{25}$~m$^{-3}$ & \\
$N_{\textrm{Mol}}$ &  Density of neutral molecules & & \\
$T_e$ & Electron temperature & & \\
$T_{e,\textrm{[eV]}}$ & Electron temperature in eV & & \\
$T_g$ & Temperature of heavy species & & \cite{papeer2014extended} \\
$T_v$ & Vibrational temperature of heavy species & & \cite{papeer2014extended} \\
\hline
\end{tabular}
\end{center}
\label{tab:parameters}
\end{table*}

Hydrodynamic as well as the spatial dynamics induced by the external DC field are not considered in the present work. 

\section{Results and discussion}

\begin{figure}[h]
\includegraphics{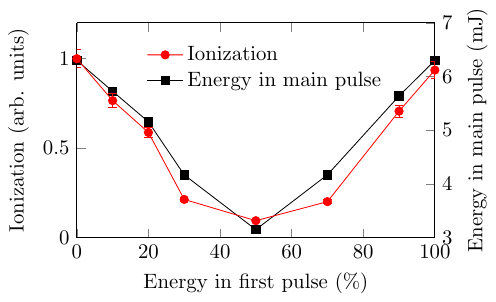} 
  \caption{Effect of energy partitioning on the ionization yield. The delay between the pulses is 4 ns and the electrode gap is 3~cm.} 
  \label{fig:Epartitioning}
 \end{figure}

\textbf{Experiments.} As displayed on Figure~\ref{fig:Epartitioning}, 
partitioning the pulse energy into a dual pulse reduces the ionization yield, regardless of the delay. In the power range of our experiment, the peak density of free electrons is governed by the power of the stronger pulse without significant contribution from the weaker one, that does not contain enough power to create a filament. 
The main action of the weaker pulse is to slighty heat up the gas.
 
At energies below 3~mJ, the beam remains below the critical power, so no filamentation occurs any more and the intensity decreases. As the ionization is highly non-linear, the electron yield drastically drops down below the detection threshold.

\begin{figure}[h]
\includegraphics{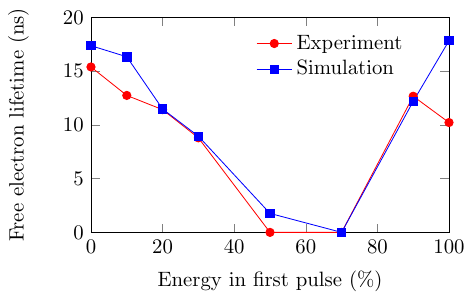} 
  \caption{Free electron lifetime when excited by a double pulse (0.5 ns delay), as a function of the relative intensity between the two pulses, in a 3~cm gap under 15~kV. }
    \label{fig:duree_Eratio}
 \end{figure}

\begin{figure}[h]
\includegraphics{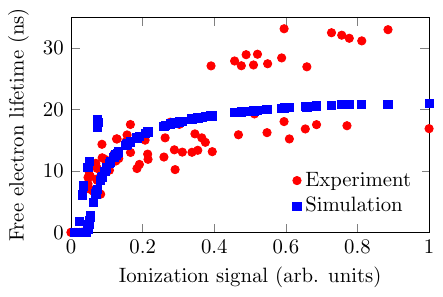} 
\caption{Free electron lifetime as a function of their initial density. The red dots are the experimental data points and the blue squares the simulation.}  \label{fig:MaxPeak_Lifetime}
\end{figure}

The effect of the pulse partitioning on the free electron lifetime (Figure~\ref{fig:duree_Eratio}) is to a large extent governed by the drop in the initial electron density, that results in hitting faster the streamer threshold. 
The asymmetry in the curve is due to the electron density threshold used to determine the free electron lifetime and to the second pulse being only 0.5~ns after the first one. At the energy levels measured here, this asymetry disapears for delays between the pulses longer than 1~ns.
Similar results are obtained over the whole range of investigated pulse delays, i.e. between 0.5 and 9.5~ns and for both electrode gaps (1~cm, 10~kV and 3~cm, 15~kV). The decrease of ionization appearing where the pulse is split equally also applies for triple and quadruple pulses, where the intensity is sufficiently low for the ionization level to remain below the detection limit. 
The photodetachment and re-ionization by the second pulse do not compensate the less efficient ionization, as evidenced on~\cref{fig:MaxPeak_Lifetime} where each point represents the data measured or simulated in one experimental condition, i.e. for a given energy ratio and delay between the pulses. The set of points covers all the energy ratio from 1:0 to 0:1 and delays from 0.5 to 9.5~ns. The bifurcation appearing close to zero ionization signal in the theoretical data is an artefact due to the absolute threshold used to determine the free electron lifetime. Monitoring the contribution of the photodetachment and re-ionization during the simulations confirm this interpretation. Furthermore, with an energy limited to a few mJ, the second pulse does not significantly heat the plasma. 

More specifically, this free electron decay is governed 
by both electron attachment to oxygen molecules and recombination of electrons. 
Attachment has a typical time of 23~ns, independent from the initial electron density as the reservoir of oxygen molecules is virtually infinite for the electron densities at play in the present work ($< 2 \times 10^{15}$~cm$^{-3}$). This attachment rate, that strongly depends on temperature, is consistent with the values of a few tens of nanoseconds put forward by Zhao et al.~\cite{ZhaoDWE1995}.

In contrast, recombination depends quadratically on the ionization yield. Its contribution increases for larger initial electron densities, and therefore reduces their lifetime. The typical recombination time amounts to 380~ns  for an initial electron density of 2~$\times$~10$^{13}$~cm$^{-3}$: under these conditions, recombination is negligible. In contrast, the recombination time drops down to 10~ns for the highest initial electron densities relevant to the present work~(1.6~$\times$~10$^{15}$~cm$^{-3}$).

This competition between recombination and attachment also allows us to estimate the absolute electron density in our experiment: an average free electron decay time of 7~ns at $1/e$ for a single pulse corresponds to  $1.6 \times 10^{15}$ electrons/cm$^3$.

\textbf{Simulations.} In turn, this calibration of the electron density is used to evaluate the incident pulse intensity $I_L$ for each pulse energy by inverting Equation~\eqref{eq:Ne} for the duration of the pulse.

Based on this input, the model reproduces well the experimental data (\cref{fig:duree_Eratio,fig:MaxPeak_Lifetime}).

\begin{figure}[h]
\begin{center}
\includegraphics{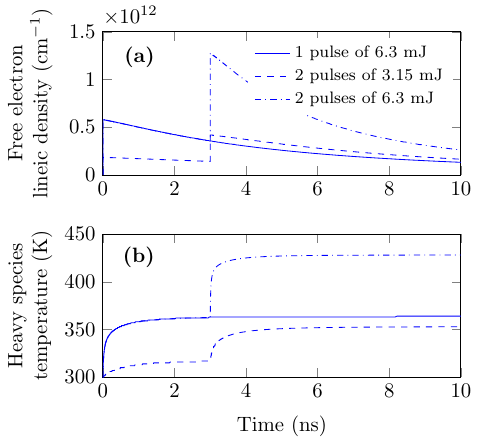} 
\end{center} 
\caption{(a) Free electron lineic density and (b) heavy species temperature reached with respect to time for one laser pulse of 6.3~mJ or two laser pulses of respectively 3.15~mJ and 6.3~mJ each.}
\label{fig:comparison_1fullE_2fullE_2halfE}
\end{figure}

Furthermore, it illustrates the effect of splitting the pulse energy in the experiments. 
\Cref{fig:comparison_1fullE_2fullE_2halfE}a compares the simulated free electron lineic density for one single filamenting 6.3~mJ pulse and two 3.15~mJ sub-pulses. As observed in the experiment, 
splitting the energy in two pulses delayed with 3~ns only decreases the available electron density. On the other hand, increasing the total energy to 12.6~mJ by sending two pulses of each 6.3~mJ increases more than twice the electron density than for the single pulse and the free electron lifetime is approximately doubled, as observed by~\cite{PolynM2011}.
\Cref{fig:comparison_1fullE_2fullE_2halfE}b shows the gas temperature along the laser beam propagation axis for the same conditions. As for the free electron density, as soon as the energy in a pulse is sufficient to create a filament, the temperature rises significantly. As the heavy species have a long vibrational relaxation time, the gas temperature stays high for a very long time and leads to an expansion of the gas. This in turn allows the free electrons to travel unperturbed on a longer distance, increasing the discharge efficiency \cite{VidalCCDLJKMPR2000,papeer2014extended}.

\begin{figure}[h]
\includegraphics{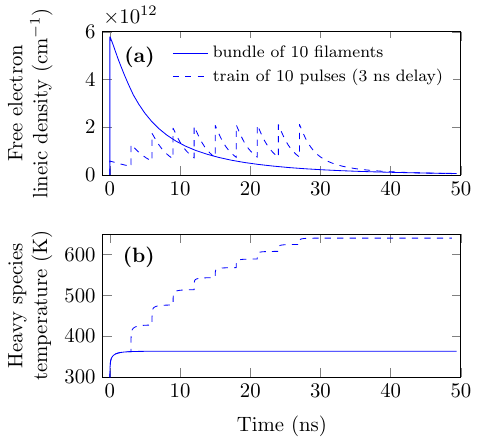} 
\caption{(a) free electron lineic density on the beam propagation axis and (b) heavy species temperature with respect to time for a bundle of 10 filaments and a train of 10 subpulses of 63~mJ total energy.}
\label{fig:10pulses_Ne}
\end{figure}

We now investigate how the partitioning effect evolves for high total energies, i.e., enough energy to create 10 filaments that we can split either in time (train of pulses) or in space (multifilament bundle). 
As an example, \cref{fig:10pulses_Ne} shows the free electron lineic density created by a bundle of ten filaments and for a train of ten filaments with 3~ns delay between them and the corresponding gas temperature increase. 
As our interest is atmospheric applications, we consider the case of a collimated or loosely focused beam. In this case, the individual filaments of the multiple filamentation pattern have similar properties, including ionization level, than a single filament~\cite{ChinHLLTABKKS2005,CouaiM2007,BergeSNKW2007}. This assumption would not hold in the case of tightly focused beams, where the ionization and the energy deposition increases~\cite{ThebeLSBC2006,point2016energy}. Such situation would require to include a detailed simulation of the propagation, which is beyond the scope of the present work. 

It is very clear from \cref{fig:10pulses_Ne}a that a bundle of filaments creates less free electrons at a given time but the free electron density is always higher than that of a single filament. The free electron lifetime is governed by the successive re-ionizations and is therefore comparable to the length of the pulse train. 
On the other hand, each subpulse creating one filament is depositing energy in the same parcel of  gas, increasing its temperature, whereas the bundle of filaments deposits much less energy per unit volume in the gas. Therefore, a train of pulses with sufficient energy in each pulse to create a filament is more favorable to trigger an electrical discharge, as the free electron density created is high and the streamer can propagate on longer distances thanks to the air rarefaction due to the heavy species temperature increase.

Note that a more precise evaluation of this effect would need to consider the cross-pulse influence via the ionization left behind by the pulses. For example, long pulses will influence the propagation of the subsequent pulses~\cite{bejot2011transition}. This is however beyond the scope of this present work.

\section{Conclusion}

As a conclusion, due to the high non-linearity of the ionization process that overrides plasma heating effects as well as photodetachment and photo-ionization, partitioning the energy of an ultrashort laser pulse  of 6.3~mJ, 450~fs pulse   into trains of two to four pulses decreases the yield in free electrons. Accordingly, the plasma lifetime during which it keeps above the typical electron density of a leader decreases.

Trains of pulses are nevertheless efficient to increase the free electron lifetime if each subpulse carries enough energy to initiate a filament. Also, as the pulse trains are very efficient to heat the heavy species, this leads to a longer electron mean free path, which should help triggering the discharge by allowing more electrons to travel along the streamer channel.

We expect that our results will help dimensioning and optimizing the pulse shape of future systems aimed at triggering high-voltage discharges and guiding them, especially over large scales.

\subsection{Acknowledgments}
This work was supported by the ERC advanced grant "Filatmo". J.-G. B. acknowledges support from the FP7 ITN network "JMAP", and M. M. from SNSF through the Marie Heim-V\"ogtlin grant.
We gratefully acknowledge fruitful discussions with Nicolas Berti, Denis Mongin and Julien Guillod, as well as experimental support by Michel Moret. 

\section{References}

\bibliography{Biblio}

\end{document}